\documentclass[twocolumn,secnumarabic,amssymb, nobibnotes, aps, prd]{revtex4-1}

\usepackage{amsmath,amssymb,bm}

\begin{document}

\title{Neutrino spin and spin-flavor oscillations in
transversal matter currents with standard and nonstandard interactions}%

\author{Pavel Pustoshny}
\email{kfrepp@gmail.com}
\affiliation{Department of Theoretical Physics, \ Moscow State University, 119991 Moscow, Russia}
\author{Alexander Studenikin}
\email{ studenik@srd.sinp.msu.ru}
\affiliation{Department of Theoretical Physics, \ Moscow State University, 119991 Moscow, Russia}
\address{Joint Institute for Nuclear Research, 141980 Dubna, Russia}

\date{Jule 6, 2018}%
\begin{abstract}
After a brief  history of two known types of neutrino mixing and oscillations, including neutrino spin and spin-flavor oscillations in the transversal magnetic field, we perform a systematic study of  a new phenomenon of neutrino spin and spin-flavor oscillations engendered by the transversal matter currents on the bases of the developed quantum treatment of the phenomenon. Possibilities for the resonance amplification of these new types of oscillations by the longitudinal matter currents and longitudinal magnetic fields are analyzed. Neutrino spin-flavor oscillations engendered by the transversal matter currents in the case of nonstandard interactions of neutrinos with background matter are also considered.
\end{abstract}
\maketitle
\tableofcontents

\section{Introduction}

Neutrino mixing and oscillations are  no doubt among of the most exciting  and intriguing phenomena of the present fundamental physics. Being introduced into the physics of elementary particles
 on the basis of sufficiently general theoretical principles more than 60 years ago \cite{Pontecorvo:1957cp}, these phenomena over the past few decades not only made it possible to obtain a solution to the problem of solar and atmospheric neutrinos, but also marked the beginning of a campaign into a new physics. There are two principal types of neutrino oscillations: flavor and spin oscillations. The former arise when there is an initial {\it inherent} mixing of the mass states of the neutrinos, the latter can occur due to the mixing of neutrinos with different polarizations when the magnetic moment of the particle interacts with an external magnetic field.

 In this paper, we consider in detail the new possibility of the appearance of spin and spin-flavor neutrino oscillations engendered by  weak interactions of neutrinos with the medium in the case when there are the transversal currents of matter. For the neutrino spin oscillations in this case there is no need neither for a neutrino nonzero  magnetic moment nor for an external magnetic field presence.

 For the first time the phenomenon of spin oscillations of neutrinos due to weak interactions with the transversal matter currents and/or the transversal polarization of matter was considered in \cite{Studenikin:2004bu} (this possibility was also mentioned in \cite{Studenikin:2004tv}). The existence of the discussed effect of neutrino spin oscillations engendered by the transversal matter current  was confirmed in a series of recent papers \cite{Cirigliano:2014aoa, Volpe:2015rla, Kartavtsev:2015eva, Dobrynina:2016rwy,Tian:2016hec} where its possible impact in astrophysics was also discussed.

Given these circumstances
there is an urgent need for a systematic and consistent presentation of the theory of this new phenomenon. Indeed, the discussed neutrino
spin (and spin-flavor) oscillations engendered
by the transversal matter current is a new type of oscillations
that have never been discussed before the publication of the paper \cite{Studenikin:2004bu}. Therefore, it would be useful to recall the main
points of the history of the neutrino oscillation phenomena studies.

The paper is organized as follows. After   a brief history of neutrino mixing and oscillations, give in Sec.  \ref{Sec_2}, we present in Sec. \ref{Sec_3} the semiclassical and then quantum treatment of the neutrino spin oscillations $\nu_{e}^L\Leftarrow (j_{\perp}) \Rightarrow \nu_{e}^R$ engendered by the transversal matter currents and derive the corresponding neutrino evolution Hamiltonian. In Sec. \ref{Sec_4} the probability of the neutrino spin oscillations $\nu_{e}^L\Leftarrow (j_{\perp}, B_{\perp}) \Rightarrow\nu_{e}^R$ in the transversal matter current and a constant magnetic field is derived. It is shown that there are possibilities for the resonance amplification of the considered spin oscillations  $\nu_{e}^L \Leftarrow (j_{\perp}) \Rightarrow \nu_{e}^R$ engendered by the transversal matter current due to: (1) the longitudinal matter currents, and (2) the longitudinal  magnetic field. In Sec. \ref{Sec_5} the  neutrino spin-flavor  oscillations  $\nu_{e}^L \Leftarrow (j_{\perp}) \Rightarrow\nu_{\mu}^R$ engendered by the  transversal matter current are considered and the resonance amplification of the corresponding probability is discussed. The neutrino spin-flavor  oscillations  $\nu_{e}^L \Leftarrow (j_{\perp}) \Rightarrow\nu_{\mu}^R$ engendered by the  transversal matter current with the nonstandard interactions
are considered in Sec. \ref{Sec_6}.

\section{A brief history of neutrino mixing and oscillations}
\label{Sec_2}

{\it{Neutrino flavor oscillations in vacuum and matter.}}

The story of the neutrino mixing and oscillations started with two papers by Bruno Pontecorvo \cite{Pontecorvo:1957cp,Pontecorvo:1957qd} where the above-mentioned effects have been discussed for the first time. In \cite{Pontecorvo:1957cp} Pontecorvo has indicated that if the neutrino charge were not conserved then the transition between a neutrino and antineutrino would become possible in vacuum. In \cite{Pontecorvo:1957qd} Pontecorvo has even directly introduced a phenomenon of neutrino mixing. He has written the following.

{\it "Neutrinos in vacuum can transform themselves into antineutrino and vice versa. This means that neutrino and antineutrino are particle mixtures. So, for example, a beam of neutral leptons from a reactor which at first consists mainly of antineutrinos will change its composition and at a certain distance $R$ from the reactor will be composed of neutrino and antineutrino
in equal quantities."}

The paper \cite{Pontecorvo:1957qd} ends with the following statement.

{\it``Under the above assumptions, effects of transformation of neutrino into antineutrino and vice versa may be unobservable in the laboratory because of large value of $R$, but will certainly occur, at least, on an astronomic scale."}

A brief history of neutrino mixing and oscillations can be found in \cite{Bilenky:2016pep}.
In 1962, just after the discovery of the second flavor neutrino, the effect of neutrino mixing was discussed in \cite{Maki:1962mu} where
the fields of the weak neutrinos $\nu_e$ and $\nu_{\mu}$ were connected with the neutrinos mass states
$\nu_1$ and $\nu_2$ by the unitary mixing matrix $U$ that can be parametrized by the mixing angle $\theta$ and
\begin{equation}
\nu_e = \nu_1 \cos \theta + \nu_2 \sin \theta, \ \ \  \nu_\mu = - \nu_1 \sin \theta + \nu_2 \cos \theta .
\end{equation}

The theory of neutrino mixing and oscillations was further developed in \cite{Gribov:1968kq, Bilenky:1975tb} with actual calculations of neutrino beam evolution. In \cite{Wolfenstein:1977ue} the effect of neutrino interaction with matter of a constant density on neutrino flavor mixing and oscillations was investigated. The existence of resonant amplification of neutrino mixing  (the MSW effect) when a neutrino flux propagates through a medium with varying density was predicted in \cite{Mikheev:1986gs}.

The tedious studies, both experimental and theoretical, over the past 60 years has been honored by the Nobel Prize of 2015 awarded to Arthur McDonald and Takaaki Kajita for the discovery of neutrino oscillations, which shows that neutrinos have mass.

{\it{Neutrino spin oscillations in magnetic fields.}}

The straightforward consequence of neutrino nonzero mass is the prediction \cite{Fujikawa:1980yx} that neutrinos can have nonzero magnetic moments. Studies of neutrino magnetic moments and the related phenomena attract a reasonable interest in literature.  The values of neutrino magnetic moments are constrained in the terrestrial laboratory experiments and in the astrophysical considerations (see, for instance, \cite{Beda:2012zz} and \cite{Raffelt:1990pj}).

Massive neutrinos participate in electromagnetic interactions. The recent review on this topic is given in
  \cite{Giunti:2014ixa} (the upgrade can be found in \cite{Studenikin:2018vnp}). One of the most important phenomena of nontrivial neutrino electromagnetic
interactions is the neutrino magnetic moment precession and the corresponding
spin oscillations in the presence of external
electromagnetic fields. The later effect has been studied in numerous papers published
during several passed decades.

Within this scope the neutrino spin oscillations \ \ \ \ \ \ \ \ $\nu^{L}\Leftrightarrow \nu^{R}$ induced by the neutrino magnetic moment interaction with the transversal magnetic field ${\bf B}_{\perp}$ was first considered in \cite{Cisneros:1970nq}. Then spin-flavor oscillations $\nu^{L}_{e}\Leftrightarrow \nu^{R}_{\mu}$ in ${\bf B}_{\perp}$ in vacuum were discussed
in \cite{Schechter:1981hw}, the importance of the matter effect was emphasized in \cite{Okun:1986hi}.
The effect of the resonant amplification of neutrino spin oscillations in ${\bf B}_{\perp}$ in the presence of matter was proposed in \cite{Akhmedov:1988uk,Lim:1987tk}, the impact of the longitudinal magnetic field ${\bf B}_{||}$ was discussed in \cite{Akhmedov:1988hd}.
The neutrino spin oscillations in the presence of constant twisting magnetic field
were considered in \cite{Vidal:1990fr, Smirnov:1991ia, Akhmedov:1993sh,
Likhachev:1990ki,Dvornikov:2007aj,Dmitriev:2015ega}.

Recently a new approach to the description of neutrino spin and spin-flavor oscillations in the presence of an arbitrary constant magnetic field have been  developed  \cite{Dmitriev:2015ega,Studenikin:2016zdx, Popov:2018seq}. Within the new approach exact quantum stationary states  are used for classification of neutrino spin states, rather than the neutrino helicity states that have been  used for this purpose within the customary approach in many published papers. Recall that the helicity states are not stationary in the presence of a magnetic field. It has been shown \cite{Popov:2018seq}, in particular, that in the presence of the transversal magnetic field for a given choice of parameters (the  energy and magnetic moments of neutrinos and strength of the magnetic field)  the amplitude of the flavor oscillations $\nu_e^L \Leftrightarrow \nu_{\mu}^L$ at the vacuum frequency is modulated by the magnetic field frequency. Similar results on the important influence of the transversal magnetic field on amplitudes of various types of neutrino oscillations were obtained earlier \cite{Kurashvili:2017zab} on the basis of the exact solution of the effective equation for neutrino evolution in the presence of a magnetic field and matter, which accounts for four neutrino species corresponding to two different flavor states with positive and negative helicities.

In \cite{Egorov:1999ah} neutrino spin oscillations were considered in the presence of  arbitrary constant electromagnetic fields $F_{\mu \nu}$. Neutrino spin oscillations in the presence of the field of circular and linearly polarized electromagnetic waves and superposition of an electromagnetic wave and constant magnetic field were considered in \cite{Lobanov:2001ar, Dvornikov:2001ez,Dvornikov:2004en}.

The more general case of neutrino spin evolution in the case when the neutrino is subjected to general types of non-derivative interactions with external scalar $s$, pseudoscalar $\pi$, vector $V_{\mu}$, axial-vector $A_{\mu}$, tensor $T_{\mu\nu}$ and pseudotensor $\Pi_{\mu\nu}$ fields was considered in \cite{Dvornikov:2002rs}. From the  general neutrino spin evolution equation, obtained in \cite{Dvornikov:2002rs}, it follows that neither scalar $s$ nor pseudoscalar $\pi$ nor vector $V_{\mu}$ fields can induce neutrino spin evolution. On the contrary, within the general consideration of neutrino spin evolution it was shown that electromagnetic (tensor) and weak (axial-vector) interactions can contribute to the neutrino spin evolution.

Recently we have considered in detail \cite{Fabbricatore:2016nec,Studenikin:2016zdx} neutrino mixing and oscillations in the arbitrary constant magnetic field that have  ${\bf B}_{\perp}$ and ${\bf B}_{||}$ nonzero components and derived an explicit expressions for the effective neutrino magnetic moments
for the flavor neutrinos in terms of the corresponding magnetic moments
introduced in the neutrino mass basis.

 \section{Neutrino spin oscillations  $\nu_{e}^L \Leftarrow (j_{\perp}) \Rightarrow \nu_{e}^R$ engendered by transversal matter currents}
\label{Sec_3}

For many years, until  2004, it was believed that a neutrino helicity precession
and the corresponding spin oscillations can be induced by the neutrino magnetic
interactions with an external electromagnetic field that provided the existence of the transversal magnetic field component ${\bf B}_{\perp}$ in the particles rest frame.
A new and very interesting possibility for neutrino spin
(and spin-flavor) oscillations engendered by the neutrino interaction with matter background was
proposed and investigated for the first time in \cite{Studenikin:2004bu}. It
was shown \cite{Studenikin:2004bu} that neutrino spin oscillations
can be induced not only by the neutrino interaction with a  magnetic field, as it was believed
before, but also by neutrino interactions with matter in the case when there
is a transversal matter current or matter polarization. This new effect has been
explicitly highlighted in \cite{Studenikin:2004bu}.

{\it ``The possible emergence of neutrino spin oscillations  owing to neutrino
interaction with matter under the condition that there exists a nonzero transverse current component or
matter polarization is the most important new effect that follows from the investigation
of neutrino-spin oscillations in Sec. IV. So far, it has been assumed that neutrino-spin oscillations
may arise only in the case where there exists a nonzero transverse magnetic field in the neutrino rest
frame."}

For historical notes reviewing studies of the discussed effect, see in \cite{Studenikin:2016iwq, Studenikin:2016ykv}.
It should be noted that the predicted effect exists regardless of the composition of the background matter transversal current and the source of its possible polarization (that can be a background magnetic field, for instance).

Note that the existence of the discussed effect of neutrino spin oscillations engendered by the transversal matter current and its possible impact in astrophysics was confirmed in a series of recent papers \cite{Cirigliano:2014aoa, Volpe:2015rla, Kartavtsev:2015eva, Dobrynina:2016rwy}. In the most recent paper \cite{Tian:2016hec} it has been pointed out that the effect of neutrino spin conversion from left-handed helicity states to right-handed helicity states in the absence of a magnetic field or a large magnetic moment (that was first predicted in \cite{Studenikin:2004bu}) would be present in a supernova environment.

\subsection{Semiclassical treatment}

 Following the discussion in \cite{Studenikin:2004bu} consider, as an example,  an electron neutrino spin precession in the case when neutrinos with the Standard Model interaction are propagating through moving and polarized matter composed of electrons (electron gas) in the presence of an electromagnetic field given by the electromagnetic-field tensor $F_{\mu \nu}=({\bf E}, {\bf B})$.
To derive the neutrino spin oscillation probability in the transversal matter current we use the generalized Bargmann-Michel-Telegdi equation that describes  the evolution of the
three-di\-men\-sio\-nal neutrino spin vector $\bf{S} $,
\begin{equation}\label{S}  {d{\bf S}
\over dt}={2\over \gamma} \Big[ {\bm S} \times ({\bm
B}_0+{\bm M}_0) \Big],
\end{equation}
where the magnetic field $\bf{B}_0$ in the neutrino rest frame is determined by the transversal
and longitudinal (with respect to the neutrino motion) magnetic and electric field components in the
laboratory frame,
\begin{equation}
{\bf  B}_0=\gamma\Big({\bf B}_{\perp} +{1 \over \gamma} {\bf
B}_{\parallel} + \sqrt{1-\gamma^{-2}} \Big[ {\bf E}_{\perp} \times
\frac{{\bm\beta}}{\beta} \Big]\Big),
\end{equation}
$\gamma = (1-\beta^2)^{-{1 \over 2}}$, $\bm{\beta}$ is the neutrino velocity.
The matter term ${\bf M}_0$ in Eq. (\ref{S}) is also composed of the transversal ${\bf  M}{_{0_{\parallel}}}$
and longitudinal  ${\bf  M}_{0_{\perp}}$ parts,
\begin{equation}
{\bf M}_0=\bf {M}{_{0_{\parallel}}}+{\bf M}_{0_{\perp}},
\label{M_0}
\end{equation}
\begin{equation}
\begin{array}{c}
\displaystyle {\bf M}_{0_{\parallel}}=\gamma{\bm\beta}{n_{0} \over
\sqrt {1- v_{e}^{2}}}\left\{ \rho^{(1)}_{e}\left(1-{{\bf v}_e
{\bm\beta} \over {1- {\gamma^{-2}}}} \right)\right\}- \\
-\displaystyle{\rho^{(2)}_{e}\over {1- {\gamma^{-2}}}} \left\{{\bm\zeta}_{e}{\bm\beta}
\sqrt{1-v^2_e}+ {\left({\bm \zeta}_{e}{{\bf v}_e}\frac{({\bm\beta}{\bf v}_e)}{{1+\sqrt{1-v^2_e}} }\right)}
\right\}, \label{M_0_parallel}
\end{array}
\end{equation}
\begin{equation}\label{M_0_perp}
\begin{array}{c}
\displaystyle {\bf M}_{0_{\perp}}=-\frac{n_{0}}{\sqrt {1-
v_{e}^{2}}}\Bigg\{ {\bf v}_{e_{\perp}}\Big(
\rho^{(1)}_{e}+\\
 +\rho^{(2)}_{e}\frac{({\bm\zeta}_{e} {\bf v}_e)} {1+\sqrt{1-v^2_e}}\Big) +\displaystyle {{\bm\zeta}_{e_{\perp}}}\rho^{(2)}_{e}\sqrt{1-v^2_e}\Bigg\}.
\end{array}
\end{equation}
Here $n_0=n_{e}\sqrt {1-v^{2}_{e}}$ is the invariant number density of
matter given in the reference frame for which the total speed of
matter is zero. The vectors ${\bf v}_e$, and ${\bm \zeta}_e \
(0\leq |{\bm \zeta}_e |^2 \leq 1)$ denote, respectively,
the speed of the reference frame in which the mean momentum of
matter (electrons) is zero, and the mean value of the polarization
vector of the background electrons in the above-mentioned
reference frame. The coefficients $\rho^{(1,2)}_e$ calculated
within the extended Standard Model supplied with $SU(2)$-singlet right-handed neutrino
$\nu_{R}$ are, respectively,  $\rho^{(1)}_e={\tilde{G}_F \over {2\sqrt{2}\mu }}$ and $\rho^{(2)}_e =-{G_F \over {2\sqrt{2}\mu}}$,
where $\tilde{G}_{F}={G}_{F}(1+4\sin^2 \theta _W).$

For neutrino evolution between two neutrino states $\nu_{e}^{L}\Leftrightarrow\nu_{e}^{R}$ in the presence of the magnetic field and moving matter we get \cite{Studenikin:2004bu} the following equation:
\begin{equation}\label{2_evol_eq}
	i\frac{d}{dt} \begin{pmatrix}\nu_{e}^{L} \\ \nu_{e}^{R} \\  \end{pmatrix}={\mu}
	\begin{pmatrix}
	{1 \over \gamma}\big|{\bf
M}_{0\parallel}+{{\bf B}}_{0\parallel}\big| & \big|{{\bf B}}_{\perp} + {1\over
\gamma}{\bf M}_{0\perp} \big|  \\
	 \big|{{\bf B}}_{\perp} + {1\over
\gamma}{\bf M}_{0\perp} \big| & -{1 \over \gamma}\mid{\bf
M}_{0\parallel}+{{\bf B}}_{0\parallel}\big|  \\
		\end{pmatrix}
	\begin{pmatrix}\nu_{e}^{L} \\ \nu_{e}^{R} \\ \end{pmatrix}.
\end{equation}
Thus, the probability of the neutrino spin oscillations in the adiabatic
approximation is given by \cite{Studenikin:2004bu}
\begin{eqnarray}\label{ver2}
P_{\nu^{L}_{e} \rightarrow \nu^{R}_{e}} (x)&=&\sin^{2} 2\theta_\textmd{eff}
\sin^{2}{\pi x \over L_\textmd{eff}}, \nonumber \\
\sin^{2} 2\theta_\textmd{eff}&=&{E^2_\textmd{eff} \over
{E^{2}_\textmd{eff}+\Delta^{2}_\textmd{eff}}}, \\
L_\textmd{eff}&=&{\pi \over
\sqrt{E^{2}_\textmd{eff}+\Delta^{2}_\textmd{eff}}}, \nonumber
\end{eqnarray}
where
\begin{eqnarray} \label{E3}
E_\textmd{eff}&=&\mu \big|{{\bf B}}_{\perp} + {1\over
\gamma}{\bf M}_{0\perp} \big|, \nonumber \\
\Delta_ \textmd{eff}&=&{\mu \over \gamma}\big|{\bf
M}_{0\parallel}+{{\bf B}}_{0\parallel} \big|.
\end{eqnarray}

Thus, it follows that even without the presence of an electromagnetic field,
${{\bf B}}_{\perp}={{\bf B}}_{0\parallel}=0$,
neutrino spin  oscillations $\nu_{e}^{L}\Leftrightarrow\nu_{e}^{R}$ can be induced in the presence of matter
when the transverse matter term ${\bf M}_{0\perp}$ is not zero.
If we neglect possible effects of matter polarization then the neutrino evolution equation (\ref{2_evol_eq}) simplifies to

\begin{equation}\label{3_evol_eq}
	i\frac{d}{dt} \begin{pmatrix}\nu_{e}^{L} \\ \nu_{e}^{R} \\  \end{pmatrix}=
\frac{\mu}{\gamma}
	\begin{pmatrix}
	M_{0\parallel} & M_{0\perp}   \\
	  M_{0\perp}  & -M_{0\parallel}  \\
		\end{pmatrix}
	\begin{pmatrix}\nu_{e}^{L} \\ \nu_{e}^{R} \\ \end{pmatrix},
\end{equation}
where
\begin{eqnarray}\label{M_0_parallel_perp}
\begin{array}{c}
\displaystyle {\bf M}_{0_{\parallel}}=\gamma{\bm\beta}\rho^{(1)}_{e} \left(1-{{\bf v}_e
{\bm\beta} \over {1- {\gamma^{-2}}}} \right){n_{0} \over
\sqrt {1- v_{e}^{2}}}, \\
 {\bf M}_{0_{\perp}}=-\rho^{(1)}_{e}{\bf v}_{e_{\perp}}\frac{n_{0}}{\sqrt {1-
v_{e}^{2}}}.
\end{array}
\end{eqnarray}
The effective mixing angle and oscillation length in the neutrino spin oscillation probability (\ref{ver2}) now are given by
\begin{equation}
\sin^{2} 2\theta_\textmd{eff}={M_{0\perp}^2\over
{M_{0\parallel}^2 +M_{0\perp}^2}}, \ \ \
L_\textmd{eff}={\pi \over {\mu M_0}}\gamma.
\end{equation}
The above considerations can be applied to other types of neutrinos and various matter compositions. It is also obvious that for neutrinos with nonzero transition magnetic moments a similar effect for spin-flavor oscillations exists under the same background conditions.

\subsection{Quantum treatment}
\label{3_2}

 Here below we continue our studies of the effect of neutrino spin evolution induced by the transversal matter currents and  develop a consistent derivation (see also \cite{Studenikin:NOW_2016,Studenikin:CORFU_2017}) of the effect based on the direct calculation of the spin evolution effective Hamiltonian in the case when a neutrino is propagating in  the transversal currents
 of matter.

 Consider two flavor neutrinos with two possible helicities $\nu_{f}= (\nu_{e}^{+}, \nu_{e}^{-}, \nu_{\mu}^{+}, \nu_{\mu}^{-})^T$ in moving matter composed of neutrons. The neutrino interaction Lagrangian reads
 \begin{eqnarray}\label{Lagr}
 {\it L}_{int}&=& -f^{\mu}\sum_{l}
 \bar{\nu}_l (x)\gamma_{\mu}\frac{1+\gamma _{5}}{2}\nu _l (x)=\nonumber \\
 &=& -f^{\mu}\sum_{i}  \bar{\nu}_i (x)\gamma_{\mu}\frac{1+\gamma _{5}}{2}\nu _i (x), \nonumber \\
 f^{\mu}&=&-\frac{G_F}{\sqrt 2}n(1,\bf v),
 \end{eqnarray}
 where $l= e, \ or \  \mu$ indicates the neutrino flavor, $i=1,2$ indicates the neutrino mass state and the matter potential $f^{\mu}$ depends on the neutron number density in the
 laboratory reference frame $n=\frac{n_0}{\sqrt{1-v^2}}$  and on the velocity of matter
 ${\bf v}=(v_1 , v_2 , v_3)$.
 Each of the flavor neutrinos is a superposition of the neutrino mass states,
\begin{equation}\label{transformations}
    \begin{array}{c}
    %\begin{tabular}{ccc}
  \nu_{e}^{\pm} =\nu_{1}^{\pm}\cos\theta+\nu_{2}^{\pm}\sin\theta,\ \ \ \ \
  \nu_{\mu}^{\pm}=-\nu_{1}^{\pm}\sin\theta+\nu_{2}^{\pm}\cos\theta.
  %\end{tabular}
\end{array}
\end{equation}
The neutrino evolution equation in the flavor basis is
\begin{equation}\label{schred_eq_fl}
  i\dfrac{d}{dt}\nu_{f}=\Big(H_{0} + \Delta H_{SM}\Big) \nu_{f},
\end{equation}
where the first term $H_{0}$ of the effective Hamiltonian determines the neutrino evolution in nonmoving matter. The second term $\Delta H_{SM}$ accounts for the effect of matter motion and it can be expressed as (see also \cite{Studenikin:NOW_2016,Studenikin:CORFU_2017})
\begin{equation}\label{delta_H}
\Delta H_{SM}=
\begin{pmatrix}
\Delta^{++}_{ee} & \Delta^{+-}_{ee} &\Delta^{++}_{e\mu} & \Delta^{+-}_{e\mu}  \\
\Delta^{-+}_{ee} & \Delta^{--}_{ee}  & \Delta^{-+}_{e\mu}  & \Delta^{--}_{e\mu}  \\
\Delta^{++}_{\mu e} & \Delta^{+-}_{\mu e} & \Delta^{++}_{\mu \mu} & \Delta^{+-}_{\mu \mu} \\
\Delta^{-+}_{\mu e} & \Delta^{--}_{\mu e} & \Delta^{-+}_{\mu \mu} & \Delta^{--}_{\mu \mu}
\end{pmatrix},
\end{equation}
where
\begin{equation}\label{HB}
\Delta^{ss'}_{kl}=\langle	{\nu_{k}^{s}}|\Delta H_{SM}|{\nu_{l}^{s'}}\rangle, \ \ \ k,l = e,\mu, \ \ s,s'=\pm.
\end{equation}
From (\ref{Lagr}) it follows that
\begin{equation}\label{delta_H_1}
\begin{array}{c}
\Delta H_{SM}=\frac{G_F}{2\sqrt 2}n \big(1+\gamma _{5}\big) {\bf v} {\bm \gamma}, \ \ \
{\bf v} {\bm \gamma}= v_1 \gamma _1 + v_2 \gamma _2 + v_3 \gamma _3 .
\end{array}
\end{equation}

In evaluation of $\Delta^{ss'}_{kl}$ we have first introduced
the neutrino flavor states $\nu_{k}^{s}$ and $\nu_{l}^{s'}$ as superpositions of
the mass states $\nu_{1,2}^{\pm}$. Then, using the exact
free neutrino mass states spinors,
\begin{equation}	 \label{wave func}
\nu_{\alpha}^{s}=C_{\alpha}\sqrt{\frac{E_{\alpha}+
m_{\alpha}}{2E_{\alpha}}}\begin{pmatrix}u^{s}_{\alpha} \\ \frac{\bm{\sigma p_{\alpha}}}{E_{\alpha}+m_{\alpha}}u^{s}_{\alpha}
\end{pmatrix}e^{i\bm{p_{\alpha} x}}, \ \ \ \alpha=1,2,
\end{equation}
where the two-component spinors $u^s_{\alpha}$,
\begin{equation}\label{u_s_plus}
	u^{s=1}_{\alpha}=\begin{pmatrix}1 \\ 0\end{pmatrix}, \ \ \ \
	u^{s=-1}_{\alpha}=\begin{pmatrix}0 \\ 1\end{pmatrix},
\end{equation}
define neutrino helicity states, we have performed calculations that are analogous to
those performed in \cite{Fabbricatore:2016nec}. The difference
in calculations is that here we consider not electromagnetic
neutrino interaction with a magnetic field but the neutrino weak
interaction with moving matter given by (\ref{delta_H}). For the
typical term $\Delta^{ss'}_{\alpha \alpha '}=
\langle	{\nu_{\alpha}^{s}}|\Delta H_{SM}|{\nu_{\alpha '}^{s'}}\rangle$, which
by fixing proper values of $\alpha, s, \alpha'$ and $s'$ can reproduce all
of the elements of the neutrino evolution Hamiltonian $\Delta H^{eff}$
that accounts for the effect of matter motion, we obtain (see also \cite{Studenikin:NOW_2016,Studenikin:CORFU_2017})
\begin{eqnarray} \label{Delta_ss_aa}
 \Delta^{ss'}_{\alpha \alpha '}&=&
\frac{G_F}{2\sqrt 2}\frac{n_0}{\sqrt {1-v^2}}\Big\{{u^{s}_{\alpha}}^{T}
\Big[ \big(1-\sigma_3\big)v_{\parallel}+\nonumber \\
&+&\big({\gamma_{\alpha \alpha '}}^{-1}\sigma_1 +i
{\widetilde{\gamma}_{\alpha  \alpha '}}^{-1}\sigma_2
\big)v_{\perp}\Big]
u^{s'}_{\alpha'}\Big\}\delta_{\alpha}^{\alpha'},
\end{eqnarray}
where $v_{\parallel}$ and $v_{\perp}$ are the
longitudinal and transversal velocities of the matter current and
\begin{eqnarray}
{\gamma_{\alpha \alpha '}}^{-1}&=&\frac{1}{2}\big(
\gamma_{\alpha}^{-1}+\gamma_{\alpha '}^{-1}\big), \nonumber \\
{\widetilde{\gamma}_{\alpha \alpha '}}^{-1}&=&\frac{1}{2}\big(
\gamma_{\alpha}^{-1}-\gamma_{\alpha '}^{-1}\big), \nonumber \\
\gamma_{\alpha}^{-1}&=&\frac{m_\alpha}{E_\alpha}.
\end{eqnarray}
Recalling expressions for the Pauli matrixes,
\begin{equation}\label{Paul_matrix}
\sigma _3 =\begin{pmatrix}
	1 & 0  \\
	  0 & -1  \\
		\end{pmatrix}, \ \ \ \ \sigma _1 =\begin{pmatrix}
	0 & 1  \\
	  1 & 0  \\
		\end{pmatrix},\ \ \ \ \sigma _2 =i\begin{pmatrix}
	0 & -1  \\
	  1 & 0  \\
		\end{pmatrix},
\end{equation}
we get (see also \cite{Studenikin:NOW_2016,Studenikin:CORFU_2017})
\begin{widetext}
\begin{eqnarray}\label{Delta_ss_aa_final}
%\begin{array}{c}
 \Delta^{ss'}_{\alpha \alpha '}&=&
\frac{G_F}{2\sqrt 2}\frac{n_0}{\sqrt {1-v^2}}\left\lbrace {u^{s}_{\alpha}}^{T}
\left[ \begin{pmatrix}
	0 & 0  \\
	  0  & 2  \\
		\end{pmatrix}v_{\parallel} \right. \right.   + \left. \left. \begin{pmatrix}
	0 & \gamma_{\alpha}^{-1}  \\
	  \gamma_{\alpha '}^{-1}  & 0  \\
		\end{pmatrix}v_{\perp}\right]
u^{s'}_{\alpha'}\right\rbrace \delta_{\alpha}^{\alpha'}.
%\end{array}
\end{eqnarray}
\end{widetext}

The obtained general expression (\ref{Delta_ss_aa_final}) can be used for investigations of various
types of neutrino spin oscillations in the transversal matter currents considered in the neutrino mass basis. It confirms our prediction \cite{Studenikin:2004bu}  that there are the effect of the neutrino
spin conversion and corresponding spin oscillations engendered by the interaction with
the transversal current of matter. It is also clear that the corresponding effect engendered
by the transversal polarization of matter can be treated in much the same way.

On the basis of Eq. (\ref{Delta_ss_aa_final}) and using the relation (\ref{transformations}) between neutrino mass $\nu _{\alpha}^{\pm}$ and flavor $\nu _{l}^{\pm}$ states it is possible to bring our considerations to observational terms and study neutrino oscillations in the flavor basis $\nu_{f}^s$. The neutrino flavor and mass states are connected by the neutrino mixing matrix,
\begin{equation}\label{U_def}
  \nu_{f}=U\nu_\alpha,
\end{equation}
for which in the considered case we have
\begin{equation}\label{U}
  U=
\begin{pmatrix}
    \cos\theta & 0 & \sin\theta & 0 \\
	0 & \cos\theta & 0 & \sin\theta \\
	-\sin\theta & 0 & \cos\theta & 0 \\
    0 & -\sin\theta & 0 & \cos\theta \\
\end{pmatrix}.
\end{equation}
The corresponding neutrino evolution equation is
%Leaving aside the oscillations due to kinetic part, which are are not the subject of our paper, we
\begin{equation}\label{schred_eq_fl}
  i\dfrac{d}{dt}\nu_{f}=H^{f}_{v}\nu_{f},
\end{equation}
where the effective Hamiltonian is given by $H^{f}_{v}=UHU^{\dag}$ and can be directly calculated using Eq. (\ref{U}).

 However, it is possible to get a general structure of the effective evolution Hamiltonian for the flavor neutrino using results of our previous studies \cite{Fabbricatore:2016nec,Studenikin:2016zdx} of neutrino oscillations in the arbitrary magnetic field ${\bm B} = {\bm B}_{\parallel}+{\bm B}_{\perp}$. For evaluation of the flavor neutrino oscillation in an arbitrary moving matter that is characterized by the current ${\bm j} = {\bm j}_{\parallel}+{\bm j}_{\perp}$ we consider results for the flavor neutrino oscillations in the magnetic field ${\bm B}$ and account for similarity of the correspondence between the neutrino magnetic moment to magnetic field interaction Hamiltonian $H_{B}$,
\begin{equation}
\label{aa}
H_{B} = -\mu_{\alpha \alpha '}
\bar{\nu}_{\alpha '}{ \bf{\Sigma} \bf{B}}\nu_{\alpha} + H.c. ,\ \ \ \ \Sigma_i=
\left(\begin{array}{cc}
                   \sigma_i & 0 \\
                   0 & \sigma_i
                 \end{array}\right),
\end{equation}
 and the neutrino to moving matter interaction Hamiltonian $H_{v}$
\begin{equation}
H_{v}=\widetilde{G} n\bar{\nu}_{\alpha '}{\bm v}{\bm \gamma}\nu_{\alpha},\ \ \ \ \
\widetilde{G}=\frac{G_F}{2\sqrt{2}}.
 \end{equation}
Here $n=\frac{n_0}{\sqrt{1-v^2}}=n_{0}\gamma_{n}$, $n_0$ is the invariant density of matter composed of neutrons.
For the flavor neutrino  evolution Hamiltonian in the magnetic field
$H_{B}^{f} = U H_{B} U^{\dag}$ we have  \cite{Fabbricatore:2016nec,Studenikin:2016zdx, footnote}
 %{\tiny
\begin{equation}\label{H_B_f}
 {H}_B^{f}=\begin{pmatrix}
 -(\frac{\mu}{\gamma})_{ee}{B_{||}}  & \mu_{ee} B_{\perp} & -(\frac{\mu}{\gamma})_{e\mu}{B_{||}}  & \mu_{e\mu} B_{\perp} \\
\mu_{ee} B_{\perp} & (\frac{\mu}{\gamma})_{ee}{B_{||}}  & \mu_{e\mu} B_{\perp} &  (\frac{\mu}{\gamma})_{e\mu}{B_{||}}  \\
-(\frac{\mu}{\gamma})_{e\mu}{B_{||}}  & \mu_{e\mu} B_{\perp} & -(\frac{\mu}{\gamma})_{\mu\mu}{B_{||}}  & \mu_{\mu\mu} B_{\perp} \\
\mu_{e\mu} B_{\perp} & (\frac{\mu}{\gamma})_{e\mu}{B_{||}}  & \mu_{\mu\mu} B_{\perp} &  (\frac{\mu}{\gamma})_{\mu\mu}{B_{||}}
\end{pmatrix},
\end{equation}

where the following notations are used
\begin{eqnarray}\label{mu_Gammas_fl}
  \nonumber \Big(\frac{\mu}{\gamma}\Big)_{ee} &=& \frac{\mu_{11}}{\gamma_{11}}\cos^2\theta+\frac{\mu_{22}}{\gamma_{22}}\sin^2\theta+\frac{\mu_{12}}{\gamma_{12}}
  \sin2\theta, \\  \Big(\frac{\mu}{\gamma}\Big)_{e\mu} &=& \frac{\mu_{12}}{\gamma_{12}}\cos2\theta+\frac{1}{2}\Big(\frac{\mu_{22}}{\gamma_{22}}-\frac{\mu_{11}}{\gamma_{11}}
  \Big)\sin2\theta ,\\
  \nonumber \Big(\frac{\mu}{\gamma}\Big)_{\mu\mu} &=& \frac{\mu_{11}}{\gamma_{11}}\sin^2\theta+\frac{\mu_{22}}{\gamma_{22}}\cos^2\theta-\frac{\mu_{12}}
  {\gamma_{12}}\sin2\theta,
\end{eqnarray}
\begin{eqnarray}\label{mu_fl}
  \mu_{ee} &=& \mu_{11}\cos^2\theta+\mu_{22}\sin^2\theta+\mu_{12}\sin2\theta, \nonumber \\
  \mu_{e\mu} &=& \mu_{12}\cos2\theta+\frac{1}{2}(\mu_{22}-\mu_{11})\sin2\theta, \\
  \mu_{\mu\mu} &=& \mu_{11}\sin^2\theta+\mu_{22}\cos^2\theta-\mu_{12}\sin2\theta .\nonumber
\end{eqnarray}

For the flavor neutrino  evolution Hamiltonian in moving matter $H_{v}^{f} = U H_{v} U^{\dag}$ we get\\
\begin{widetext}
\begin{equation}\label{H_v}
H^f_v=n\tilde{G}\left( \begin{matrix}
0& (\frac{\eta}{\gamma})_{ee}v_{\perp} &0 &(\frac{\eta}{\gamma})_{e\mu}v_{\perp}\\
(\frac{\eta}{\gamma})_{ee}v_{\perp} & 2\eta_{ee}(1-v_{\parallel}) &(\frac{\eta}{\gamma})_{e\mu}v_{\perp}& \eta_{e\mu}\\
0& (\frac{\eta}{\gamma})_{e\mu}v_{\perp} &0 &(\frac{\eta}{\gamma})_{\mu \mu}v_{\perp}\\
(\frac{\eta}{\gamma})_{e\mu}v_{\perp} & \eta_{e\mu} &(\frac{\eta}{\gamma})_{\mu\mu}v_{\perp}& 2\eta_{\mu\mu}(1-v_{\parallel})\\
\end{matrix} \right),
\end{equation}
\end{widetext}
where $\frac{\eta}{\gamma}$ and $\eta$ are given by
\begin{equation}\label{eta}
\Big(\frac{\eta}{\gamma}\Big)_{ee}= \Big(\frac{\mu}{\gamma}\Big)_{ee_{\mid\mu_{11}=\mu_{22}=1, \ \mu_{12}=0}}=
\frac{\cos^2\theta}{\gamma_{11}}+\frac{\sin^2\theta}{\gamma_{22}},
 \end{equation}
\begin{equation}\label{eta}
\Big(\frac{\eta}{\gamma}\Big)_{\mu\mu}= \Big(\frac{\mu}{\gamma}\Big)_{\mu\mu_{\mid\mu_{11}=\mu_{22}=1, \ \mu_{12}=0}}=
\frac{\sin^2\theta}{\gamma_{11}}+\frac{\cos^2\theta}{\gamma_{22}},
 \end{equation}
\begin{equation}\label{eta}
\Big(\frac{\eta}{\gamma}\Big)_{e\mu}= \Big(\frac{\mu}{\gamma}\Big)_{e\mu_{\mid\mu_{11}=\mu_{22}=1, \ \mu_{12}=0}}=
\frac{\sin 2\theta}{\tilde{\gamma}_{21}},
 \end{equation}
\begin{equation}\label{eta_2}
\eta_{ee}=\mu_{{ee}_{\mid\mu_{11}=\mu_{22}=1, \ \ \mu_{12}=0}}=1,
\end{equation}
\begin{equation}\label{eta_2}
\eta_{\mu\mu}=\mu_{{\mu\mu}_{\mid\mu_{11}=\mu_{22}=1, \ \ \mu_{12}=0}}=1,
\end{equation}
\begin{equation}\label{eta_2}
\eta_{e\mu}=\mu_{{e\mu}_{\mid\mu_{11}=\mu_{22}=1, \ \mu_{12}=0}}= 0
\end{equation}
if one sets $\mu_{11}=\mu_{22}=1, \ \mu_{12}=0$ in Eqs. (\ref{mu_Gammas_fl}) and (\ref{mu_fl}).

From the above consideration it follows that neutrino spin oscillations can be engendered by the magnetic moment interactions with the transversal magnetic field $\bm {B}_{\perp}$ as well as by the neutrino weak interactions
with the transversal matter current ${\bm j}_{\perp}$.

\section{Probability of neutrino spin oscillations $\nu_{e}^L\Leftarrow (j_{\perp},B_\perp) \Rightarrow \nu_{e}^R$ engendered by transversal matter current and magnetic field}
\label{Sec_4}

Consider the initial neutrino state
$\nu_{e}^L$ moving in the background with the magnetic field ${\bm B} = {\bm B}_{\parallel}+{\bm B}_{\perp}$ and nonzero matter current ${\bm j} = {\bm j}_{\parallel}+{\bm j}_{\perp}$.One of the possible modes of neutrino transitions with the change of helicity is $\nu_{e}^L\Leftarrow (j_{\perp}, B_{\perp}) \Rightarrow \nu_{e}^R$. The corresponding oscillations are governed by the evolution equation
\begin{widetext}
\begin{eqnarray}\label{nu_L_nu_R}
	i\frac{d}{dt} \begin{pmatrix}\nu^L_{e} \\ \nu^R_{e} \\  \end{pmatrix}=
	 \left( \begin{matrix}
	(\frac{\mu}{\gamma})_{ee}{B_{||}}+\eta_{ee}{\widetilde{G}}n(1-{\bm v}{\bm \beta})\\
	 \mu_{ee}B_{\perp}+ (\frac{\eta}{\gamma})_{ee}{\widetilde{G}}nv_{\perp}
	 \end{matrix} \right.
	 \left. \begin{matrix} \mu_{ee}B_{\perp} + (\frac{\eta}{\gamma})_{ee}{\widetilde{G}}nv_{\perp}  \\
	  - (\frac{\mu}{\gamma})_{ee}{B_{||}} -\eta_{ee}{\widetilde{G}}n(1-{\bm v}{\bm \beta}) \end{matrix} \right)
	\begin{pmatrix}\nu^L_{e} \\ \nu^R_{e} \\ \end{pmatrix}.
\end{eqnarray}
\end{widetext}
For the oscillation $\nu^L_{e} \Leftarrow (j_{\perp}, B_{\perp}) \Rightarrow \nu^R_{e}$ probability we get
\begin{eqnarray}\label{prob_oscillations}
P_{\nu^L_{e} \rightarrow \nu^R_{e}} (x)&=&{E^2_\textmd{eff} \over
{E^{2}_\textmd{eff}+\Delta^{2}_\textmd{eff}}}
\sin^{2}{\pi x \over L_\textmd{eff}},\nonumber\\
 L_\textmd{eff}&=&{\pi \over
\sqrt{E^{2}_\textmd{eff}+\Delta^{2}_\textmd{eff}}},
\end{eqnarray}
where
\begin{eqnarray}\label{E3}
E_{eff}= \Big|\mu_{ee}\bm{B}_{\perp} + \Big(\frac{\eta}{\gamma}\Big)_{ee}{\widetilde{G}}n\bm{v}_{\perp}  \Big|, \nonumber \\
\Delta_{eff}= \Big|\Big(\frac{\mu}{\gamma}\Big)_{ee}{\bm{B}_{||}}+\eta_{ee}{\widetilde{G}}n(1-{\bm v}{\bm \beta}){\bm {\beta}} \Big|.
\end{eqnarray}

In the next section we estimate values of the corresponding parameters that characterize the properties of neutrinos, the background matter and the magnetic field for which neutrino spin oscillations $\nu_{e}^L\Leftarrow (j_{\perp}) \Rightarrow\nu_{e}^R$, engendered by neutrino weak interactions with the transversal matter current, can proceed with significant probability.

\subsection{Resonance amplification of neutrino spin oscillations $\nu_{e}^L\Leftarrow (j_{\perp}) \Rightarrow\nu_{e}^R$ by longitudinal matter current}

We are interested in the situation
when the amplitude of oscillations $\sin^{2} 2\theta_\textmd{eff}$ in (\ref{prob_oscillations}) is not small and we use the criterion based on the demand that
\begin{equation}\label{criterion}
\sin^{2} 2\theta_\textmd{eff}={E^2_\textmd{eff} \over
{E^{2}_\textmd{eff}+\Delta^{2}_\textmd{eff}}}\geq \frac{1}{2},
\end{equation}
which is provided by the condition $E_{eff}\geq\Delta_{eff}$.

At first we consider the case when the effect of magnetic field is negligible and thus we have
\begin{equation}\label{E3_B_zero}
E_{eff}= \Big|\Big(\frac{\eta}{\gamma}\Big)_{ee}{\widetilde{G}}n\bm{v}_{\perp}  \Big|, \ \
\Delta_{eff}= \Big| {\widetilde{G}}n(1-{\bm v}{\bm \beta}){\bm {\beta}} \Big|
\end{equation}
and the oscillation length is given by
\begin{equation}\label{osc_L_B_zero}
L_{eff}=\frac{\pi}{(\frac{\eta}{\gamma})_{ee}{\widetilde{G}}n {v}_{\perp}}.
\end{equation}
From the condition $E_{eff}\geq\Delta_{eff}$ it follows that
\begin{equation}\label{cond_2}
\Big(\frac{\eta}{\gamma}\Big)_{ee}{v}_{\perp} \geq (1-{\bm v}{\bm \beta}).
\end{equation}
In the further evaluations we suppose that $\Delta m = m_2 - m_1 \ll m_1, \ m_2$, and introduce the neutrino effective gamma-factor $\gamma _{\nu}$
\begin{equation}\label{est_gamma}
\frac{1}{\gamma_{\nu}}=\frac{1}{\gamma_{11}}\sim\frac{1}{\gamma_{22}}.
\end{equation}
Then the condition (\ref{cond_2}) reduces to
\begin{equation}\label{cond_3}
\frac{v_\perp}{\gamma_\nu} \geq (1-{\bm v}{\bm \beta}).
\end{equation}
Assuming  neutrino masses  $m_1, \ m_2 \sim 0.1$  eV, for a typical neutrino energy $p^{\nu}_0 \sim 10$ MeV we find ${\gamma_{\nu}}\sim 10^{7}$. Consider the case when neutrinos are more relativistic particles than the background matter neutrons ($\gamma _{\nu} \gg \gamma _n$), then from (\ref{cond_3}) we get
\begin{equation}\label{cond_4}
\frac{1}{\gamma_\nu} \geq \frac{1}{\gamma _{n}^2}.
\end{equation}
The latter condition can be valid for ultrarelativistic background matter with $\gamma_n\geq {\gamma_\nu}^{1/2}\sim 3\times 10^3$. At the same time the oscillation length $L_{eff}$ given by
(\ref{osc_L_B_zero}) can be $L_{eff}\sim 50$ km in the case $n\sim 10^{37}$ cm$^{3}$ and $\gamma_n\sim 3\times 10^3$.

\subsection{Resonance amplification of neutrino spin oscillations $\nu_{e}^L\Leftarrow (j_{\perp}) \Rightarrow\nu_{e}^R$ by longitudinal  magnetic field}
\label{Sec_4_2}

The presence of the longitudinal magnetic field ${\bm{B}_{||}}$ can also have an important impact on the criterion (\ref{criterion}). In the previous consideration the diminishing value of $\Delta_{eff}$ is attained by the vanishing value of $(1-{\bm v}{\bm \beta})$. Now we consider reduction of the term
${\widetilde{G}}n(1-{\bm v}{\bm \beta})$ in $\Delta_{eff}$ given by (\ref{E3}) due to the contribution of $(\frac{\mu}{\gamma})_{ee}{\bm{B}_{||}}$. This possibility can be realized when ${\bm{B}_{||}}{\bm {\beta}} = -1$. An environment we are considering can be realized by models of short gamma-ray bursts (sGRB)
(see \cite{Grigoriev:2017wff} and \cite{Perego:2014fma}). Consider the neutrino $\nu_e$ escaping the central neutron star with inclination given by an angle $\alpha$ from the plane of the accretion disk. Then this neutrino propagates through the toroidal  bulk of very dense matter that rotates with the angular velocity of about $\omega = 10 ^3$ s$^{-1}$ around the axis that is perpendicular to the accretion disk. The diameter of the perpendicular cut of the toroidal bulk of matter is about $d\sim 20$ km and the distance from the centre of this cut to the centre of the neutron star is also about  $D\sim 20$ km. The transversal velocity of matter can be estimated accordingly $v_{\perp}=\omega D= 0.067$ that corresponds to $\gamma _{n} =1.002$.

Suppose that the direction of the neutrino propagation is characterized by $\sin \alpha \sim \frac{1}{2}$. If there is a magnetic field $B$ perpendicular to the accretion disk then there is the longitudinal field in respect to the neutrino propagation ${B}_{||}=B \sin \alpha \sim \frac{1}{2} B$.

In the straightforward analysis we are particularly interested in the conversion $\nu_{e}^L\Leftarrow (j_{\perp}) \Rightarrow\nu_{e}^R$ engendered by interactions with the transversal matter current ${\bm j}_{\perp}$. Therefore we omit the possible effect of the neutrino magnetic moment interaction with the transversal magnetic field $\mu_{ee}\bm{B}_{\perp}$ in (\ref{E3}) and get
\begin{equation}
E_{eff}= \Big(\frac{\eta}{\gamma}\Big)_{ee}{\widetilde{G}}n {v}_{\perp}= \frac{\cos^2\theta}{\gamma_{11}}{\widetilde{G}}n {v}_{\perp} \approx
{\widetilde{G}}n_0\frac{\gamma_n}{\gamma_\nu}{v}_{\perp}.
\end{equation}
In the considered geometry ${\bm v}{\bm \beta}=0$ and for  $\Delta_{eff}$ we use the relation

\begin{eqnarray}
\Delta_{eff} &=& \Big|(\frac{\mu}{\gamma}\Big)_{ee}{\bm{B}_{||}} +\eta_{ee}{\widetilde{G}}n{\bm {\beta}}\Big|=\nonumber \\ &=&
\Big|{\bm{B}_{||}}\Big(
\frac{\mu_{11}}{\gamma_{11}}\cos^2\theta+\frac{\mu_{22}}{\gamma_{22}}
\sin^2\theta \Big)+{\widetilde{G}}n{\bm {\beta}}\Big|\nonumber
  \\ &\approx& \Big|\frac{\mu_{11}}{\gamma_{\nu}}B_{||}-{\widetilde{G}}n_0\gamma_n\Big|.
\end{eqnarray}
From the demand  $E_{eff}>\Delta_{eff} $ we get
\begin{eqnarray}
\Big| \frac{\mu_{11}B_{||}}{{\widetilde{G}}n_0\gamma_n}-\gamma_{\nu}\Big|\leq1.
\end{eqnarray}
Thus, the criterion (\ref{criterion}) is fulfilled if the the longitudinal magnetic field is
\begin{equation}\label{B_cr}
B_{||}=B^{cr}_{||}\sim \gamma_n \gamma_{\nu}\frac{\widetilde{G}n_0}{\mu_{11}}.
\end{equation}
From the obtained estimation of the critical strength $B^{cr}_{||}$ of the longitudinal magnetic field $B_{||}$
follows its important dependence on the matter density $n_0$. If one takes an estimation for the neutrino magnetic moment $\mu_{11} \sim 3\times 10^{-11} \mu_{B}$ and $\gamma_{\nu} = 2\times 10^{7}$ then in  the case of very low matter density $n_0 \sim 10^{23}$ cm$^{-3}$ the critical field is $B^{cr}_{||}\sim 8\times 10^{-3} B_0$, where $B_0=\frac{m_e ^2}{e_0}= 4.41 \times 10^{13}$ Gauss. For higher densities $n_0 \sim 10^{36}$ cm$^{-3}$ we get $B^{cr}_{||}\sim 10^{24}$ Gauss that is of the order of the critical field strength $B_W=\frac{m_W^2}{e_0}$ where a
component of the W-field becomes tachyonic (see \cite{Nielsen:2013daa} and references therein).

From these estimations it would seem that in order to get a reasonable reduction of the term ${\widetilde{G}}n(1-{\bm v}{\bm \beta})$ in $\Delta_{eff}$ given by (\ref{E3}) due to the contribution of $(\frac{\mu}{\gamma})_{ee}{\bm{B}_{||}}$ the matter density should not be too high to avoid a demand for extremely strong magnetic fields. However, one should also consider the scale of the effective oscillation length whose value in the case of the resonance $\Delta_{eff}\rightarrow 0$ is given by (\ref{osc_L_B_zero}). Even for the case $n_0 \sim 10^{36}$ cm$^{-3}$ the effective length is still extremely large $L_{eff}\sim 300$ km, which makes a hypothetically interesting possibility of the resonant amplification of spin oscillations due to the longitudinal component of the magnetic field unattainable.

\section{Neutrino spin-flavor  oscillations  $\nu_{e}^L\Leftarrow (j_{\perp}) \Rightarrow\nu_{\mu}^R$ engendered by transversal matter current}
\label{Sec_5}

Here below we consider another interesting consequence of neutrino standard interaction with the transversal matter current that produces the neutrino spin precession and oscillations accompanying in addition with  the change of the neutrino flavor. These effects are similar to the neutrino spin-flavor oscillations in the transversal magnetic field which, as it was shown in \cite {Akhmedov:1988uk,Lim:1987tk}, can be amplified by the resonance in the presence of matter.

For the considered case of $\nu_{e}^L\Leftarrow (j_{{\perp} B_{\perp}}) \Rightarrow\nu_{\mu}^R$ from (\ref{H_B_f}) and (\ref{H_v}) we get the following  neutrino evolution equation
\begin{widetext}
\begin{eqnarray}
	i\frac{d}{dt} \begin{pmatrix}\nu^L_{e} \\ \nu^R_{\mu} \\  \end{pmatrix}=
	\left( \begin{matrix}
	 -\Delta M+(\frac{\mu}{\gamma})_{ee}{B_{||}}+{\widetilde{G}}n(1-{\bm v}{\bm \beta}) \\
	  \mu_{e\mu}B_{\perp}+ (\frac{\eta}{\gamma})_{e\mu}{\widetilde{G}}nv_{\perp} \end{matrix} \right.
	 \left. \begin{matrix} \mu_{e\mu}B_{\perp} + (\frac{\eta}{\gamma})_{e\mu}{\widetilde{G}}nv_{\perp}  \\
	 \Delta M- (\frac{\mu}{\gamma})_{\mu\mu}{B_{||}} -{\widetilde{G}}n(1-{\bm v}{\bm \beta})
		\end{matrix} \right)
	\begin{pmatrix}\nu^L_{e} \\ \nu^R_{\mu} \\ \end{pmatrix},
\end{eqnarray}
\end{widetext}
where
\begin{equation}
\Delta M=\frac{\Delta m ^2 \cos 2\theta}{4 p^{\nu}_0 },
\end{equation}
and $p^{\nu}_0$ is neutrino energy. For the oscillation $\nu^L_{e} \Leftarrow (j_{\perp}) \Rightarrow \nu^R_{\mu}$ probability we get

\begin{eqnarray}\label{prob_oscillations_1}
P_{\nu^L_{e} \rightarrow \nu^R_{\mu}} (x)&=&\sin^{2} 2\theta_\textmd{eff}
\sin^{2}{\pi x \over L_\textmd{eff}},\nonumber \\
 \sin^{2} 2\theta_\textmd{eff}&=&{E^2_\textmd{eff} \over
{E^{2}_\textmd{eff}+\Delta^{2}_\textmd{eff}}}, \nonumber \\
L_\textmd{eff}&=&{\pi \over
\sqrt{E^{2}_\textmd{eff}+\Delta^{2}_\textmd{eff}}},
\end{eqnarray}
where
\begin{eqnarray}\label{E_1_delta_1}
E_{eff}= \Big|\mu_{e\mu}{B}_{\perp} + \Big(\frac{\eta}{\gamma}\Big)_{e\mu}{\widetilde{G}}n{v}_{\perp}  \Big|, \nonumber \\
\Delta_{eff}
=\Big|\Delta M-\frac{1}{2}\Big(\frac{\mu_{11}}{\gamma_{11}} +\frac{\mu_{22}}{\gamma_{22}}\Big){B}_{\parallel} -{\widetilde{G}}n(1-{\bm v}{\bm \beta})  \Big|.
\end{eqnarray}
From (\ref{E_1_delta_1}) it follows that the value of $\Delta_{eff}$ can be diminished by both the neutrino interaction with the longitudinal magnetic field and also by the effect of interaction with matter.
\\

\section{Resonance amplification of neutrino spin-flavor oscillations $\nu_{e}^L\Leftarrow (j_{\perp}) \Rightarrow\nu_{\mu}^R$}

We are again  interested in the situation
when the amplitude of oscillations $\sin^{2} 2\theta_\textmd{eff}$ in (\ref{prob_oscillations_1}) is not small and we use the criterion based on the demand that
\begin{equation}\label{criterion_1}
\sin^{2} 2\theta_\textmd{eff}={E^2_\textmd{eff} \over
{E^{2}_\textmd{eff}+\Delta^{2}_\textmd{eff}}}\geq \frac{1}{2},
\end{equation}
which is provided by the condition $E_{eff}\geq\Delta_{eff}$.  Thus we get the condition
\begin{eqnarray}
\Big|\mu_{e\mu}{B}_{\perp} + \Big(\frac{\eta}{\gamma}\Big)_{e\mu}{\widetilde{G}}n{v}_{\perp}\Big|\geq \nonumber \\
\geq \Big|\Delta M-\frac{1}{2}\Big(\frac{\mu_{11}}{\gamma_{11}} +\frac{\mu_{22}}{\gamma_{22}}\Big){B}_{\parallel} -{\widetilde{G}}n(1-{\bm v}{\bm \beta})   \Big|.
\end{eqnarray}
Consider the case when the effect of the magnetic field is negligible, thus we get
\begin{equation}
\Big| \Big(\frac{\eta}{\gamma}\Big)_{e\mu}{\widetilde{G}}n{v}_{\perp}\Big|\geq\Big|\Delta M-{\widetilde{G}}n(1-{\bm v}{\bm \beta})   \Big|.
\end{equation}
The effective oscillation length read
\begin{equation}\label{1_osc_L_B_zero}
L_{eff}=\frac{\pi}{\Big(\frac{\eta}{\gamma}\Big)_{e\mu}{\widetilde{G}}n {v}_{\perp}}.
\end{equation}

In the further evaluations we use the approximation
\begin{equation}
\Big(\frac{\eta}{\gamma}\Big)_{e\mu}\approx\frac{\sin 2 \theta}{\gamma_{\nu}},
\end{equation}
and get the resonance condition in the form
\begin{equation}
 \frac{{\widetilde{G}}n{v}_{\perp}}{\gamma_{\nu}}\sin 2 \theta+{\widetilde{G}}n(1-{\bm v}{\bm \beta}) \geq\Delta M .
\end{equation}
In the case ${v}_{\parallel}=0$ we get
\begin{equation}
 \frac{{\widetilde{G}}nv_{\perp}}{\gamma_{\nu}}\sin 2 \theta +{\widetilde{G}}n \approx {\widetilde{G}}n.
\end{equation}
Finally, the criterion (\ref{criterion_1}) is fulfilled when the following condition is valid
\begin{equation}\label{condition_1}
\widetilde{G}n\geq\Delta M.
\end{equation}

Consider again an environment peculiar to models of short gamma-ray bursts discussed in Sec. \ref{Sec_4_2}
(see also \cite{Grigoriev:2017wff} and \cite{Perego:2014fma}). The mass squared difference and mixing angle are taken from the solar neutrino measurements, $\Delta m^2=7.37\times 10^{-5}$  eV$^2$, $\sin ^2 \theta=0.297$ ($\cos 2 \theta=0.406$) \cite{Patrignani:2016xqp}. Consider neutrino with energy $p^{\nu}_0=10^6$ eV and  moving matter characterized by $\gamma _{n} =1.002$. Thus, we get
\begin{equation}
\Delta M=0.75\times 10^{-11} \text{eV}.
\end{equation}
Accounting for the estimation
\begin{equation}
\widetilde{G}=\frac{G_F}{2\sqrt{2}}=0.4\times 10^{-23} \ \ \text{eV}^{-2},
\end{equation}
from the criterion (\ref{condition_1}) we get the quite reasonable condition on the density of neutrons
\begin{equation}\label{n_0}
n_0 \geq\frac{\Delta M}{\widetilde{G}}=10^{12}\ eV^3\approx 10^{26}\  \text{cm}^{-3}.
\end{equation}
The corresponding oscillation length is approximately
\begin{equation}\label{1_osc_L_B_zero}
L_{eff}=\frac{\pi}{(\frac{\eta}{\gamma})_{e\mu}{\widetilde{G}}n {v}_{\perp}}\approx 5\times 10^{11} \ \text{km}.
\end{equation}
The oscillation length can be within the scale of short gamma-ray bursts discussed in Sec. \ref{Sec_4_2}
\begin{equation}\label{1_osc_L_B_zero}
L_{eff}\approx 10 \ \text{km}
\end{equation}
if the matter density equals $n_0 \approx 5\times 10^{36}\  \text{cm}^{-3}$.

\section{Neutrino spin-flavor  oscillations  $\nu_{e}^L\Leftarrow (j_{\perp}) \Rightarrow\nu_{\mu}^R$ engendered by nonstandard interactions}
\label{Sec_6}

Quite recently \cite{Stapleford:2016jgz} it has been shown that nonstandard interactions (NSI) of neutrinos  with matter \cite{Ohlsson:2012kf, Miranda:2015dra, Farzan:2017xzy} can significantly alter neutrino flavor evolution in supernovae with the potential to impact explosion dynamics, nucleosynthesis, and the neutrinos signal.

Obviously, neutrino spin oscillations can be also engendered by NSI of neutrinos with the transversal matter currents.

The effect of nonstandard interactions is usually parametrized by a set of matrices
that introduce new contributions to the matter potential of neutrinos. The strength of this new potential is
 dependent on constituent of the matter and can also contain off-diagonal contributions
known as flavor changing neutral currents.
Thus, the neutrino spin-flavor oscillations, for instance  $\nu_{e}^L\Leftarrow (j_{\perp}) \Rightarrow\nu_{\mu}^R$,  can be engendered by the nonstandard interactions with the transversal matter current.

From the Lagrangian of NSI \cite{Ohlsson:2012kf, Miranda:2015dra, Farzan:2017xzy}
\begin{eqnarray}
-L_{NSI}^{eff}=\varepsilon^{fP}_{\alpha \beta}2\sqrt{2}G_{F}(\bar{\nu}_{\alpha}
\gamma_{\rho}L\nu_{\beta})(\bar{f}\gamma^{\rho}Pf), \nonumber \\
 L,R=(1\pm \gamma^5)/2
\end{eqnarray}
(where $f=u,d,e$ and $\alpha$, $\beta=e$, $\mu$)  for the medium with only neutrons one gets
\begin{eqnarray}\label{L_NSI}
-L_{NSI}^{eff}&=& -f^{\mu}\sum_{\alpha, \beta}(\varepsilon^{uL}_{\alpha \beta}+2\varepsilon^{dL}_{\alpha \beta})\bar{\nu}_{\alpha} (x)\gamma_{\mu}\frac{1+\gamma _{5}}{2}\nu _{\beta} (x)= \nonumber \\
&=&-f^{\mu}\sum_{\alpha, \beta}\varepsilon^n_{\alpha \beta}\bar{\nu}_{\alpha} (x)\gamma_{\mu}\frac{1+\gamma _{5}}{2}\nu _{\beta} (x),
\end{eqnarray}
where the notation is introduced
\begin{equation}
\varepsilon^n_{\alpha \beta}=\varepsilon^{uL}_{\alpha \beta}+2\varepsilon^{dL}_{\alpha \beta}.
\end{equation}
The neutrino evolution equation  with standard and nonstandard interactions in the flavor basis is
\begin{equation}\label{schred_eq_fl}
  i\dfrac{d}{dt}\nu_{f}=\Big(H_{0} + \Delta H_{SM} + \Delta H_{NSI}\Big) \nu_{f}.
  \end{equation}

Accounting for the neutrino mixing, contributions to the sum (\ref{L_NSI}) can be expressed in terms of the neutrino mass states. For the characteristic contributions in (\ref{L_NSI}) we get the following expressions ($\Gamma_\mu=\gamma_{\mu}\frac{1+\gamma _{5}}{2}$):
\begin{eqnarray}
\label{1}\varepsilon_{ee}^n \bar{\nu}_e^{s} \Gamma^{\lambda}  \nu_e^{s'} &=& \varepsilon_{ee}^n (\bar{\nu}_1^{s} \Gamma^{\lambda} \nu_1^{s'} \cos^2 \theta \bar{\nu}_2^{s} \Gamma^{\lambda} \nu_2^{s'} \sin^2 \theta  +\nonumber \\
&+&\frac{1}{2} (\bar{\nu}_1^{s} \Gamma^{\lambda} \nu_2^{s'} + \bar{\nu}_2^{s} \Gamma^{\lambda} \nu_1^{s'}) \sin 2 \theta ),\\
\label{2}\varepsilon_{\mu \mu}^n \bar{\nu}_{\mu}^{s} \Gamma^{\lambda}  \nu_{\mu}^{s'} &=& \varepsilon_{\mu \mu}^n (\bar{\nu}_1^{s} \Gamma^{\lambda} \nu_1^{s'} \sin^2 \theta+\bar{\nu}_2^{s} \Gamma^{\lambda} \nu_2^{s'} \cos^2 \theta  -\nonumber \\
&-&\frac{1}{2} (\bar{\nu}_1^{s} \Gamma^{\lambda} \nu_2^{s'} + \bar{\nu}_2^{s} \Gamma^{\lambda} \nu_1^{s'}) \sin 2 \theta ),\\
\label{3}\varepsilon_{e \mu}^n \bar{\nu}_{e}^{s} \Gamma^{\lambda}  \nu_{\mu}^{s'} &=& \varepsilon_{e \mu}^n (-\bar{\nu}_2^{s} \Gamma^{\lambda} \nu_1^{s'} \sin^2 \theta+\bar{\nu}_1^{s} \Gamma^{\lambda} \nu_2^{s'} \cos^2 \theta  -\nonumber \\
&-&\frac{1}{2} (\bar{\nu}_1^{s} \Gamma^{\lambda} \nu_1^{s'} - \bar{\nu}_2^{s} \Gamma^{\lambda} \nu_2^{s'}) \sin 2 \theta ),\\
\label{4}\varepsilon_{\mu e}^n \bar{\nu}_{\mu}^{s} \Gamma^{\lambda}  \nu_{e}^{s'} &=& \varepsilon_{ \mu e}^n (\bar{\nu}_2^{s} \Gamma^{\lambda} \nu_1^{s'} \cos^2 \theta-\bar{\nu}_1^{s} \Gamma^{\lambda} \nu_2^{s'} \sin^2 \theta  -\nonumber \\
&-&\frac{1}{2} (\bar{\nu}_1^{s} \Gamma^{\lambda} \nu_1^{s'} - \bar{\nu}_2^{s} \Gamma^{\lambda} \nu_2^{s'}) \sin 2 \theta ).
\end{eqnarray}

 The matrix elements of the NSI contribution to the evolution Hamiltonian can be calculated accounting for (\ref{1})-(\ref{4}) and, as it is described in Sec. \ref{3_2}, in  the neutrino mass basis using the vacuum neutrino wave functions given by (\ref{wave func}). Note that the evolution equation in the neutrino mass basis, accounting only for the NSI, is
 \begin{equation}\label{schred_eq_fl}
  i\dfrac{d}{dt}\nu^s_{\alpha}=\Big(H^m_{0} + \Delta H^m_{NSI} \Big) \nu^s_{\alpha}.
    \end{equation}
 Here $H^m_{0}$ is the neutrino evolution Hamiltonian for the mass states in vacuum, ${\alpha}= 1,2$ and $s=\pm 1$, and
\begin{widetext}
\begin{eqnarray}\label{H_nsi}
\Delta H^m_{NSI}=\tilde{G}n
\times \left( \begin{matrix}
0 & \frac{v_{\perp}}{\gamma_{11}}e_{11}& 0 &  \frac{v_{\perp}}{\gamma_{11}}e_{12}\\
\frac{v_{\perp}}{\gamma_{11}}e_{11} &2(1-v_{\parallel}) e_{11} & \frac{v_{\perp}}{\gamma_{22}}e_{12}& 2(1-v_{\parallel})e_{12} \\
0 & \frac{v_{\perp}}{\gamma_{22}}e_{12} & 0 &  \frac{v_{\perp}}{\gamma_{22}}e_{22} \\
\frac{v_{\perp}}{\gamma_{11}}e_{12}& 2(1-v_{\parallel})e_{12} &  \frac{v_{\perp}}{\gamma_{22}}e_{22} & 2(1-v_{\parallel})e_{22} \\
\end{matrix} \right),
\end{eqnarray}
\end{widetext}
where
\begin{equation}
e_{11}=\varepsilon_{ee}^n\cos^2 \theta+\varepsilon_{\mu \mu}^n\sin^2 \theta-\varepsilon_{e \mu}^n\sin 2 \theta,
\end{equation}
\begin{equation}
e_{12}=\frac{1}{2}(\varepsilon_{ee}^n-\varepsilon_{\mu \mu}^n)\sin 2 \theta+\varepsilon_{e \mu}^n\cos 2\theta,
\end{equation}
\begin{equation}
e_{22}=\varepsilon_{ee}^n\sin^2 \theta+\varepsilon_{\mu \mu}^n\cos^2 \theta+\varepsilon_{ e\mu}^n\sin 2 \theta.
\end{equation}

Summing up three contributions to the neutrino Hamiltonian in
(\ref{schred_eq_fl}) for the neutrino spin-flavor oscillations   $\nu_{e}^L\Leftarrow (j_{\perp}) \Rightarrow\nu_{\mu}^R$, we get
\begin{widetext}
\begin{eqnarray}\label{general_evolution}
	i\frac{d}{dt} \begin{pmatrix}\nu^L_{e} \\ \nu^R_{\mu} \\  \end{pmatrix}
	=\left( \begin{matrix}
	 -\Delta M+(\frac{\mu}{\gamma})_{ee}{B_{||}}+2{\widetilde{G}}n(1-{\bm v}{\bm \beta})(1+\tilde{\varepsilon}_{ee})\\
	 \mu_{e\mu}B_{\perp}+ (\frac{\eta}{\gamma})_{e\mu}{\widetilde{G}}nv_{\perp} (1+\tilde{\varepsilon}_{e\mu})
	 \end{matrix} \right.  \ \ \  \ \
	 \left. \begin{matrix}
	  \mu_{e\mu}B_{\perp} + (\frac{\eta}{\gamma})_{e\mu}{\widetilde{G}}nv_{\perp}(1+\tilde{\varepsilon}_{e\mu})  \\
	  \Delta M- (\frac{\mu}{\gamma})_{\mu\mu}{B_{||}}  \\
		\end{matrix} \right)
	\begin{pmatrix}\nu^L_{e} \\ \nu^R_{\mu} \\ \end{pmatrix},
\end{eqnarray}
\end{widetext}
where
\begin{equation}
\tilde{\varepsilon}_{ee}=h_{22} \cos^2 \theta+h_{44} \sin^2 \theta+h_{24} \sin 2 \theta,
\end{equation}
\begin{equation}
\tilde{\varepsilon}_{e\mu}=-\frac{h_{14}}{\gamma_{11}} \sin^2 \theta+\frac{h_{23}}{\gamma_{22}} \cos^2 \theta+\frac{1}{2}\Big(\frac{h_{34}}{\gamma_{22}}-\frac{h_{12}}{\gamma_{11}}\Big) \sin 2 \theta.
\end{equation}
Note that in the neutrino spin-flavor evolution equation (\ref{general_evolution}) the standard and nonstandard interactions with matter and also possible influence of the magnetic field are accounted for.

For the oscillation $\nu^L_{e} \Leftarrow (j_{\perp}) \Rightarrow \nu^R_{\mu}$ probability in the considered case we get
\begin{eqnarray}\label{prob_oscillations_2}
P_{\nu^L_{e} \rightarrow \nu^R_{e}} (x)&=&\sin^{2} 2\theta_\textmd{eff}
\sin^{2}{\pi x \over L_\textmd{eff}},\nonumber\\
 \sin^{2} 2\theta_\textmd{eff}&=&{E^2_\textmd{eff} \over
{E^{2}_\textmd{eff}+\Delta^{2}_\textmd{eff}}},
\end{eqnarray}
\begin{equation}
L_\textmd{eff}={\pi \over
\sqrt{E^{2}_\textmd{eff}+\Delta^{2}_\textmd{eff}}}\nonumber ,
\end{equation}
where
\begin{equation}
E_{eff}= \Big|\mu_{e\mu}{B}_{\perp} + \Big(\frac{\eta}{\gamma}\Big)_{e\mu}{\widetilde{G}}n{v}_{\perp}(1+\tilde{\varepsilon}_{e\mu})   \Big|,
\end{equation}
and
\begin{equation}
\Delta_{eff}= \bigg|\Delta M-\frac{1}{2}\bigg(\Big(\frac{\mu}{\gamma}\Big)_{ee}+
\Big(\frac{\mu}{\gamma}\Big)_{\mu\mu}\bigg){B}_{\parallel} -{\widetilde{G}}n(1-{\bm v}{\bm \beta})(1+\tilde{\varepsilon}_{ee})   \bigg|.
\end{equation}

The amplitude of oscillations $\sin^{2} 2\theta_\textmd{eff}$ in (\ref{prob_oscillations_2}) is not small, when
\begin{equation}\label{criterion}
\sin^{2} 2\theta_\textmd{eff}={E^2_\textmd{eff} \over
{E^{2}_\textmd{eff}+\Delta^{2}_\textmd{eff}}}\geq \frac{1}{2}.
\end{equation}
Thus, we arrive to the condition
\begin{widetext}
\begin{eqnarray}
\bigg|\mu_{e\mu}{B}_{\perp} + \Big(\frac{\eta}{\gamma}\Big)_{e\mu}{\widetilde{G}}n{v}_{\perp}(1+\tilde{\varepsilon}_{e\mu}) \bigg|\geq \bigg|\Delta M-\frac{1}{2}\bigg(\Big(\frac{\mu}{\gamma}\Big)_{ee}+
\Big(\frac{\mu}{\gamma}\Big)_{\mu\mu}\bigg){B}_{\parallel} -{\widetilde{G}}n(1-{\bm v}{\bm \beta})(1+\tilde{\varepsilon}_{ee})   \bigg|.
\end{eqnarray}
\end{widetext}
We are in particular  interested in the effect of the NSI, thus neglecting the influence of the magnetic field we get
\begin{equation}
\Big| \Big(\frac{\eta}{\gamma}\Big)_{e\mu}{\widetilde{G}}n{v}_{\perp}(1+\tilde{\varepsilon}_{e\mu}) \Big|\geq\Big|\Delta M-{\widetilde{G}}n(1-{\bm v}{\bm \beta}) (1+\tilde{\varepsilon}_{ee})   \Big|,
\end{equation}
Using the approximation
\begin{equation}
\Big(\frac{\eta}{\gamma}\Big)_{e\mu}\approx\frac{\sin 2 \theta}{\gamma_{\nu}},
\end{equation}
and for the resonance condition for  $\nu_{e}^L\Leftarrow (j_{\perp}) \Rightarrow\nu_{\mu}^R$ we get
\begin{equation}\label{res_10}
 \frac{{\widetilde{G}}n{v}_{\perp}}{\gamma_{\nu}}(1+\tilde{\varepsilon}_{e\mu}) \sin 2 \theta+{\widetilde{G}}n(1-{\bm v}{\bm \beta})(1+\tilde{\varepsilon}_{ee}) \geq \Delta M.
\end{equation}
In the case ${v}_{\parallel}=0$,
\begin{equation}
 \frac{{\widetilde{G}}nv_{\perp}}{\gamma_{\nu}}(1+\tilde{\varepsilon}_{e\mu}) \sin 2 \theta +{\widetilde{G}}n (1+\tilde{\varepsilon}_{ee}) \approx {\widetilde{G}}n(1+\tilde{\varepsilon}_{ee}) ,
\end{equation}
and from (\ref{res_10}) we get
\begin{equation}
n \geq\frac{\Delta M}{{\widetilde{G}}(1+\tilde{\varepsilon}_{ee})}.
\end{equation}

For definiteness we use in our estimations the values of nonstandard parameters from \cite{Miranda:2015dra}:

$$\varepsilon^{uL}_{ee}=\varepsilon^{dL}_{ee}=0.3$$

$$\varepsilon^{uL}_{\mu\mu}=\varepsilon^{dL}_{\mu\mu}=0.005$$

$$\varepsilon^{uL}_{e\mu}=\varepsilon^{dL}_{e\mu}=0.023$$
 thus
$$\tilde{\varepsilon}_{ee}=0.6$$

Finally, for the  mass squared difference and mixing angle taken from the solar neutrino measurements, $\Delta m^2=7.37\times 10^{-5} \  \text{eV}^2$, $\sin ^2 \theta=0.297$, the neutrino  energy $p^{\nu}_0=10^6  \text{eV}$ and  moving matter characterized by $\gamma _{n} =1.002$ for the matter density we get
\begin{equation}\label{n_0_NSI}
n_0 \geq 0.625 \times 10^{26} \text{cm}^{-3}.
\end{equation}
The conclusion is that the account for the NSI can soften the demand on the density of the transversal matter current needed for the resonance amplification of the neutrino spin-flavor oscillations $\nu^L_{e} \Leftarrow (j_{\perp}) \Rightarrow \nu^R_{\mu}$.

\section{Conclusions}

In this paper we develop the quantum treatment of the effect of the neutrino spin $\nu_{e}^L\Leftarrow (j_{\perp}) \Rightarrow\nu_{e}^R$ and spin-flavor $\nu_{e}^L\Leftarrow (j_{\perp}) \Rightarrow\nu_{\mu}^R$ oscillations engendered by the transversal matter current that was predicted
in \cite{Studenikin:2004bu} on the basis of the semiclassical treatment of the neutrino spin evolution in the background matter. For definiteness, matter composed of neutrons is considered.

Several particular cases of the neutrino spin oscillations resonance amplifications are considered. It is shown, in particular, that the resonance in the probability of the neutrino spin oscillations $\nu_{e}^L\Leftarrow (j_{\perp}) \Rightarrow\nu_{e}^R$ can be produced by the longitudinal component of the ultrarelativistic background matter current  with $\gamma_n\geq {\gamma_\nu}^{1/2}\sim 3\times 10^3$.

The resonance amplification of the probability of the neutrino spin oscillations $\nu_{e}^L\Leftarrow (j_{\perp}) \Rightarrow\nu_{e}^R$ by the longitudinal  magnetic field has been also considered for an astrophysical environment that can be realized by models of short gamma-ray bursts (sGRB)
(see \cite{Grigoriev:2017wff} and \cite{Perego:2014fma}).

The novel effect of the neutrino {\it spin-flavor}  oscillations  $\nu_{e}^L\Leftarrow (j_{\perp}) \Rightarrow\nu_{\mu}^R$ engendered by the transversal matter current
have been also considered  and the resonance amplification of its probability has been considered for different values of the matter density $n_0 \sim 10^{26} - 10^{37}\  \text{cm}^{-3}$.

We also consider for the first time the  neutrino spin-flavor  oscillations  $\nu_{e}^L\Leftarrow (j_{\perp}) \Rightarrow\nu_{\mu}^R$ engendered by the transversal matter current with the nonstandard interactions. The oscillation probability has been derived and an estimation for the resonance matter density has been obtained with use of the realistic strengths of the neutrino NSI.

It is supposed  throughout the paper that neutrino oscillations proceed under the validity of
the adiabaticity condition, i.e. under the influence of constant or slowly varying magnetic fields and matter current densities. The opposite case (nonadiabatic regime in neutrino oscillations) requires special treatment.

For the general neutrino evolution equation,

\begin{equation}\label{gen_evol_eq}
	i\frac{d}{dt} \begin{pmatrix}\nu_{i} \\ \nu_{j} \\  \end{pmatrix}=
	\begin{pmatrix}
	H_{ii} & H_{ij}   \\
	  H_{ji}  & H_{jj}  \\
		\end{pmatrix}
	\begin{pmatrix}\nu_{i} \\ \nu_{j} \\ \end{pmatrix},
\end{equation}
the adiabaticity condition
can be represented (see \cite{Likhachev:1990ki}) in the form
\begin{widetext}
\begin{equation}\label{adiab_cond}
	\Big|(H_{jj}-H_{ii})\frac{\partial }{\partial x}(H_{ij}+H_{ji})-
(H_{ij}+H_{ji})\frac{\partial }{\partial x}(H_{jj}-H_{ii})\Big|\ll \Big[(H_{jj}-H_{ii})^2+
 (H_{ij}+H_{ji})^2\Big]^{\frac{3}{2}}.
\end{equation}
\end{widetext}
This condition can be rewritten in a more compact form,
\begin{equation}\label{1_adiab_cond}
	\Big|\Delta_{eff}\frac{\partial E_{eff}}{\partial x}-
E_{eff}\frac{\partial \Delta_{eff}}{\partial x}\Big|\ll 4 \Big[\Delta_{eff}^2+
 E_{eff}^2\Big]^{\frac{3}{2}},
\end{equation}
which, with the appropriate choice of $E_{eff}$ and $\Delta_{eff}$ values,
� can be used to analyze the situation in several specific cases discussed above.
The equation (\ref{1_adiab_cond}) clearly shows that the adiabaticity condition imposes
restrictions on changes in the strength of a magnetic field, matter density and velocity
along the direction of the neutrino propagation.

Consider, in particular, neutrino spin-flavor oscillations
$\nu_{e}^L\Leftarrow (j_{\perp}) \Rightarrow\nu_{\mu}^R$ in an environment peculiar to models of short gamma-ray bursts \cite{Grigoriev:2017wff,Perego:2014fma}
discussed in Sec. \ref{Sec_4_2}. In accordance with the model in question,
the matter density of a rotating bulk may vary in the range from $\sim 10^{38} \ \text{cm} ^3$ to
$\sim 10^{33} \ \text{cm} ^3$ at the characteristic distance of about $20$ km.
Accounting to the linear dependence the matter velocity $v$ on the neutrino traveled
distance in a rotating environment
and following the discussion on validity of the adiabaticity condition in case of
a magnetized neutron star \cite{Likhachev:1990ki}, we conclude that this condition is
fulfilled along the most neutrino travel distance.  The adiabacity condition can most
probably be violated in a quite narrow outer layer of the rotating bulk $\delta x \ll d$
which in turn is much thinner the oscillation length $L_{eff}$ given by (\ref{1_osc_L_B_zero}).
Note that in general the
origin and strengths of compact astrophysical objects magnetic fields are quite controversial
and there are various models of magnetic fields discussed in the literature
%  of there are  models that deal
%with quasi-constant magnetic fields
(see, for instance \cite{Ardeljan:2004fq, Potekhin:2015qsa, Spruit:2007bt}).

The developed quantum theory of the neutrino spin $\nu_{e}^L\Leftarrow (j_{\perp}) \Rightarrow\nu_{e}^R$ and spin-flavor $\nu_{e}^L\Leftarrow (j_{\perp}) \Rightarrow\nu_{\mu}^R$ oscillations engendered by the transversal matter current and the performed studies of different possibilities for the resonance amplification of the oscillations probabilities provide the  conclusion that these phenomena might have important consequences for generation and propagation of neutrino fluxes in extreme astrophysical interments, in particular, peculiar for supernovae. In the performed above studies we consider Dirac neutrinos, however the case of Majorana neutrinos can be treated in a quite similar way. In this concern, it is interesting to recall the statement \cite{deGouvea:2012hg, deGouvea:2013zp} that future high-precision observations of supernova fluxes, for instance, in the JUNO  experiment \cite{An:2015jdp},
may reveal the effect of the resonate amplification of Majorana neutrino spin-flavor oscillations in magnetic fields due to the neutrino collective effect. From the performed above studies it follows that the neutrino spin-flavor oscillations engendered by interactions with the transversal environment matter current, including possible neutrino self-interactions, can provide much more important effects on the observed neutrino currents in future large-volume scintillator detectors than the corresponding effect from the neutrino transition moment interaction with magnetic fields.

\section*{Acknowledgements}
The authors are thankful to Konstantin Kouzakov, Artem Popov and Konstantin Stankevich for useful discussions. This work is supported by the Russian Basic Research Foundation
Grants No. 16-02-01023 and No. 17-52-53133.

\end{document}